\documentclass[sigconf]{acmart}
\usepackage{algorithmic}
\usepackage{algorithm}
\usepackage{graphicx}
\usepackage{pgfplots}

\AtBeginDocument{%
  \providecommand\BibTeX{{%
    \normalfont B\kern-0.5em{\scshape i\kern-0.25em b}\kern-0.8em\TeX}}}

\copyrightyear{2023}
\acmYear{2023}
\setcopyright{acmlicensed}\acmConference[SecTL '23]{Secure and Trustworthy Deep Learning Systems }{July 10--14, 2023}{Melbourne, VIC, Australia}
\acmBooktitle{Secure and Trustworthy Deep Learning Systems (SecTL '23), July 10--14, 2023, Melbourne, VIC, Australia}
\acmPrice{15.00}
\acmDOI{10.1145/3591197.3591307}
\acmISBN{979-8-4007-0181-8/23/07}
\begin{document}

\title{Energy-Latency Attacks to On-Device Neural Networks via Sponge Poisoning}

\author{Zijian Wang}
\email{zwan0324@student.monash.edu}
\affiliation{%
  \institution{Monash University}
  \city{Melbourne}
  \state{Victoria}
  \country{Australia}
  \postcode{3168}
}
\author{Shuo Huang}
\email{Shuo.huang1@monash.edu}
\affiliation{
 \institution{Northwestern Polytechnical University}
 \city{Xi'an}
 \country{China}
 \postcode{710129}
  \institution{, Monash University}
  \city{Melbourne}
  \state{Victoria}
  \country{Australia}
  \postcode{3168}
}
\author{Yujin Huang}
\email{Yujin.Huang@monash.edu}
\affiliation{
  \institution{Monash University}
  \city{Melbourne}
  \state{Victoria}
  \country{Australia}
  \postcode{3168}
}
\author{Helei Cui}
\email{chl@nwpu.edu.cn}
\affiliation{
 \institution{Northwestern Polytechnical University}
 \city{Xi'an}
 \country{China}
 \postcode{710129}
}


\renewcommand{\shortauthors}{Zijian Wang}

\renewcommand\footnotetextcopyrightpermission[1]{}
\begin{abstract}    
In recent years, on-device deep learning has gained attention as a means of developing affordable deep learning applications for mobile devices. However, on-device models are constrained by limited energy and computation resources. In the mean time, a poisoning attack known as sponge poisoning has been developed.This attack involves feeding the model with poisoned examples to increase the energy consumption during inference. As previous work is focusing on server hardware accelerators, in this work, we extend the sponge poisoning attack to an on-device scenario to evaluate the vulnerability of mobile device processors.  We present an on-device sponge poisoning attack pipeline to simulate the streaming and consistent inference scenario to bridge the knowledge gap in the on-device setting. Our exclusive experimental analysis with processors and on-device networks shows that sponge poisoning attacks can effectively pollute the modern processor with its built-in accelerator. We analyze the impact of different factors in the sponge poisoning algorithm and highlight the need for improved defense mechanisms to prevent such attacks on on-device deep learning applications.
\end{abstract}




\maketitle


\section{Introduction}

Recent years have witnessed the impressive success of deep neural networks in a wide range of tasks from the fields of computer vision(CV)\cite{CV2021Junyi} and natural language process(NLP)\cite{NLP2020survey}.
Concurrently, with the advancement of hardware and compression techniques in mobile devices, the offline inference of lightweight DNNs has become feasible\cite{chen2020mobilesurvey}, enabling locally deployed inferences that eliminate the risk of information leakage and high latency caused by internet connection.  More companies take on-device deployment of DNN as a more promising trend to offer more diverse and secure services to customers like personalized text prediction\cite{xu2018deeptype}, personal item recommendation\cite{han2021deeprec} and image classification\cite{howard2019mobilenetv3}. However, limited resources remain a major challenge for practical and runtime mobile environments, where a massive and consistent model inference can cause high energy consumption and device overheating, resulting in excessive latency and decision-making failure, especially in time-sensitive tasks. \par

Recently, some works have shown that attacks on DNNs can increase inference consumption and time, exacerbating the average model energy consumption and draining power. A massive and consistent model inference that commonly appeared in streaming and dynamic tasks can still cause high energy consumption and device overheating. It may further result in excessive latency in decision-making failure, especially in time-sensitive tasks and makes the whole ruining environment shut down. For example, high latency and overheating of intelligent sensors in self-driving cars may threat human life.\par 
This concern are recently verified by some works in deep learning models, such attack can increase the inference consumption and time. As for the modern computation accelerators that boost training speed is genuinely designed to reduce overall computation cost in average computing cases by skipping insignificant computation operations. Examples can be manipulated towards worst-case operation cost when conducting inference, namely "sponge attack"\cite{shumailov2021spongeexample}. By crafting examples towards the direction of reducing insignificant computations, perturbed examples can occupy more training resources without losing any accuracy.  \par

Moreover, for hardwares that are embedded zero-skipping strategy to optimize inference cost like ASIC, a new model poisoning method, "sponge poisoning"\cite{cina2022spongepoisoning}, are developed to slow down the general inference by feeding model with toxic examples in the training phrase. It aggravates average model energy consumption in target model by reducing data sparsity in the training period and drain the power out with with normal inference examples.
Even both of these works performed in the PC and achieved significant success in deteriorating inference cost. They still encounter the limitations that can not be directly applied in on-device inference in several folds.
\begin{itemize}
    \item \textbf{Implausible Attack Scenario}: The threat is only considered in cloud service providers which is not concretely energy and resource limited attack setting. The attack test scheme is stationary and neglect the streaming data inference in real world.
    \item \textbf{Limited Victim Models and Hardwares}: The hardware and victim model can not be directly applied to mobile devices as testings are all performed in computer end. In previous work, Sponge Example has tested several variants of ResNet and DenseNet\cite{huang2017densenet}, but only 1 sample from MobileNetV2 \cite{sandler2018mobilenetv2}. And the Sponge Poisoning Attack is only applied to ResNet and VGG \cite{simonyan2014vggnet}.
    \item \textbf{Incomparable Evaluation Setting and Metrics}: The construction of the experiment and metrics are also not in line with the situation of on-device machine learning. The authors of Sponge Example mainly tested the energy consumption and latency of three cases: desktop CPU, GPU, and an individually developed ASIC simulator. The experiment of Sponge Poisoning Attack directly reuses the former ASIC simulator as the energy consumption benchmark. These settings are not comparable to on-device implementation that uses mobile tailored chips and hardware accelerators
\end{itemize}

To address this gap, we aim to conduct sponge poisoning in the mobile platform and verify the necessity of altering sponge poisoning in commercial practice, which, to the best of our knowledge, is the first paper targeting sponge poisoning in mobile devices. We perform controlled sponge training on widely used MobileNetV2 and MobileNetV3 with modern chipsets to demonstrate a significant increase in energy consumption in the inference time with clear examples. Our experiments suggest that modern mobile device processors with state-of-art accelerators may be more vulnerable to sponge poisoning than older versions without accelerators.
In this paper, we summarize the contribution as follows:
\begin{itemize}
    \item We first adapt the sponge attack to the on-device scenario. The sponge poisoning is used to comprise the accelerators and leads to as much energy overhead as possible
    \item We build the pipeline and prototype of testing the consumption in the mobile model
    \item We empirically verified the threat of sponge poisoning in terms of CPU with its built-in accelerators.
    \item Our result implies the counter-intuitive phenomenon that hardwares with more advanced accelerators for deep learning computation may have higher risk and be more vulnerable towards energy-latency related attack.
    \item Once the paper is accepted, we will open-source our verification process for the community.
\end{itemize}

\section{Background}
\subsection{Advsarial attack and data poisoning in DNN}
DNN is notorious for its interpretability and reliability. The inner training of deep learning consists of layers of linear and non-linear units which makes feature engineering an automatic process. Without human interference, the resulting high-level abstractions learned by the model may not be easily distinguishable from random linear combinations\cite{szegedy2013intriguing}, particularly after the discovery of adversarial examples\cite{goodfellow2014explaining,huang2021robustness}, where small, imperceptible perturbations can cause catastrophic performance degradation in the entire system. Due to the powerful learning ability of DNN, models tend to over-fit the features they have learned in a latent way where the determinate features are not ideal for human perception. \par
To defend the adversarial attacks, One common approach to defend against adversarial attacks is to add adversarial examples into the training phase, so the model can identify such perturbations and avoid being misled. However, attackers can also leverage this methodology to develop data poisoning attacks\cite{8jagielski2018manipulating} on DNNs, where they feed maliciously crafted examples into the training process to degrade the model's performance significantly during inference.\par
Based on targets of attack, the poisoning attacks can be further categorized into 3 scenarios\cite{cina2022Poisoning}. Indiscriminate poisoning degrades one part of the inference stage in general, \cite{solans2021fairness} shows the possibility of triggering discrimination in algorithmic judgement with poisoned examples. Targeted poisoning aims to cause targeted misclassification. It leverages complexity in the vector space to add information of certain classes to another, thus confusing the model with one class to another\cite{Shafahi2018poisonfrog}. Third, is a backdoor attack which makes a backdoor for certain perturbations, the perturbed example will result in model failure whereas a clear example will perform normally for compromised systems\cite {huang2023training}.
Besides, this threat model usually requests an untrusted training environment or untrusted data resources, especially fitting into the setting of federated learning\cite{tolpegin2020data} and online learning \cite{wang2018data} which requires decentralized data and incremental streaming training.

\subsection{Accelerating DNN with Data Sparsity}
To speed up deep learning training, one common attempt is to introduce sparsity into the system to skip the insignificant computation \cite{guo2020accelerating}\cite{qin2020sigma}.
Data sparsity is usually used to describe the number of zero-valued elements in a vector.
In DNN, sparsity is considered as the ratio of zeros in the network system\cite{wen2016learningSparsity}.
Sparsity in DNN can be divided into two categories:
  \begin{itemize}
      \item Static Sparsity: Where the zero-valued elements remain constant against different inputs. It always appears to be parameters of DNN, such as weights.
      \item Dynamic Sparsity: Where the zero-valued elements might be changed against various inputs. Such as the sparsity in activations. 
  \end{itemize}
Static sparsity is normally used to prune the model with insignificant weight, based on the indicators like the magnitude of signals\cite{zhu2017prune} or percentage information in channels\cite{xingyu2018sparseCNN}. For the dynamic sparsity, one main reason is the prevalent usage of Relu units\cite{agarap2018deep}.

\begin{equation}
    Relu(x) = max(0,x) 
\end{equation} 

The Relu function shows excellent characteristics in dealing with negative numbers and has a stable gradient signal throughout the training which makes it popular in modern deep-learning models. By using Rule, networks like autoencoders\cite{33laina2016deeper},\cite{34dosovitskiy2015flownet}, generative adversarial networks\cite{36goodfellow2020generative} can have sparsity over 90\% over its parameters which can be leveraged to skip operations.


\subsection{Hardware Accelerators with Sparsity}

In recent years, the use of hardware accelerators has become commonplace in the field of deep learning to perform matrix-level computation. Apple employs neural engines in mobile phones to achieve higher energy efficiency \cite{22cvml2017device}, while Google has introduced the use of TPUs in its data centres \cite{21jouppi2017datacenter}. Facebook and Microsoft also have similar techniques \cite{23hazelwood2018applied}\cite{24chung2018serving}. Among the variety of hardware accelerators, sparsity-based methods have received increasing   attention\cite{26parashar2017scnn}\cite{29nurvitadhi2016accelerating}.\par
In practical applications, there are mainly two approaches to improve performance through weights and activation sparsity. One way called multiply-and-accumulates (MAC) computations can gate and skip the zero-valued weight in the hardware level\cite {29nurvitadhi2016accelerating} to save the training process. Another approach focuses on optimizing computation in sparse graph and reusing existing data, and nodes in a flexible way, which increase the reuse rate of non-sparse data and improves the efficiency of the whole system \cite{41chen2019eyeriss}\cite{42zhao2019automatic}\cite{43han2016deep}.

\subsection{Attacks on Energy}\label{Attacks on Energy}
Considering the high training overhead of modern DNNs, cloud subscription services by ML-as-a-Service providers seem to be an affordable option for a small company. This situation naturally leads to concerns about targeted and centralized attacks. Some researchers even argue that over-subscription can easily expose data centres to power attacks \cite{44li2016power}\cite{45somani2016ddos}\cite{46xu2014power}\cite{47xu2015measurement}. If an attacker can maliciously trigger power spikes on the service provider's server clusters, there is a high possibility of overloading the system and eventually causing service outages \cite{44li2016power}\cite{48palmieri2015energy}. For mobile devices, fast battery consumption can cause users inconvenience in daily life. In
\begin{figure*}
        \centering
        \includegraphics[width=\textwidth]{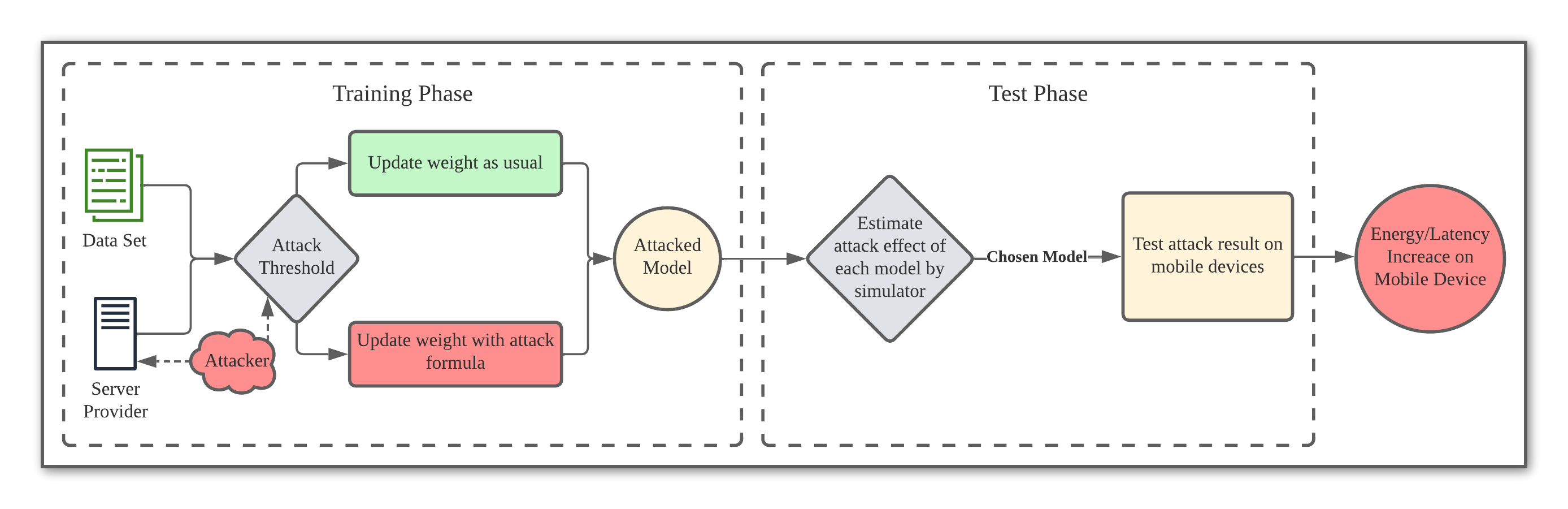}
        \caption{Energy-Latency Attacks' Pipeline}
        \Description{}
        \label{fig:pipeline}
\end{figure*}
addition, the hardware temperature increasing caused by high power consumption can also lead to serious consequences. For example, a temperature difference of 15°C can sometimes lead to a sharp increase in the failure rate \cite{49anderson2003more}. For many mobile devices, overheating may trigger severe throttling. The overall power consumption tends to increase\cite{50efraim2012energy}. And the synthesis results in a non-linear dependence between power consumption and latency.

\subsection{Security of Machine Learning}
Adapted from the concepts of conventional security, "CIA", which stands for confidentiality, integrity, and availability, exclusive researches have been done in recent decades
Attacks on DNN focused on confidentiality and integrity, such as target model stealing\cite{53juuti2019prada} that uses either direct parameter exploration, model imitation or membership inference\cite{3shokri2017membership} that mines the user information from vectored input or gradient signal. 

Unlike the iconic attack among network security, denial of service \cite{51bellardo2003802}\cite{52ferguson1998network}, which targets the availability of the system, attacks have been neglected in the field of DNNs. Until adversarial attacks \cite{kurakin2018adversarial} expose serious security flaws in DNNs, the sparking attention to its safety-critical applications in self-driving cars, unmanned aerial vehicles, and healthcare deployment have arisen.

\subsection{Energy-latency related Attack}\label{Sponge Poisoning}
There are two representative energy attack method in research community. The Sponge Example Attack\cite{cina2022spongepoisoning} and Sponge Poisoning \cite{shumailov2021spongeexample}.The Sponge Poisoning Attack is tightly built upon the Sponge Example. 
The Sponge examples introduce new threats to the availability of DNNs based on energy consumption and latency. By inputting carefully crafted and optimized data, DNNs deployed on hardware accelerators are forced to have higher energy consumption and latency when processing tasks than usual. However, in the current stand, the process of finding proper sponge examples is computational and time-consuming. \par
Inspired by data poisoning, based on the increasingly common situation of ML-as-a-Service, Sponge Poisoning Attack puts forward the hypothesis of malicious service providers. It proposes an attack method that directly interferes with the model parameters during the training process, which can deliver a trained model with increasing energy and latency but an unnoticeable accuracy drop. 
However, both works are not addressing this issue to energy and latency constraint on-device scenario where the entire training setting is inherently different from  cloud services.\par

\subsection{Mobile Machine Learning Platforms}\label{Mobile Phone Platforms}
While running many state-of-the-art deep learning models on mobile phones is initially a challenge as it is often not optimized for mobile inference. This situation changed in 2015 with the launch of TensorFlow Mobile as the demand for mobile devices became more popular and soon ushered in TensorFlow lite and Android Neural Network API (NNAPI) in 2017 \cite{huangy2021robustness,huang2022smart,sang2023beyond}. After the technique explosion in this area, it has been commonplace for the mobile device to equip an AI-specialized acceleration module. \par
The past research has summed up the development of mobile deep learning well and generally divided it into several generations till 2019 \cite{ignatov2019aismartphones}. The selection of subsequent test platforms in this paper also refers to the development route of mobile chips to a certain extent:
\begin{itemize}
    \item There are no AI accelerations, but can still support special SDKs or GPU-based libraries which can slightly accelerate machine learning inference. Such as Qualcomm SoCs with Hexagon 682 DSP.
    \item Supporting NNAPI from 2017, which gives them the ability to provide accelerations for some specific types of models. In a sense, this is the first generation of mobile chips that truly support hardware-level AI acceleration. Such as Snapdragon 845, 710; Hi Silicon Kirin 970; Samsung Exynos 9810, 9610; MediaTek Helio P70. 
    \item Providing hardware accelerations for all model types based on the previous generation. Such as Snapdragon 855, 730; Hi Silicon Kirin 980; Samsung Exynos 9825, 9820; MediaTek Helio P90. 
\end{itemize}

Prior to the publication of this paper, mobile processors underwent a significant period of development and advancement. However, it has been demonstrated in some instances that mobile hardware accelerators share similar application ideas with their desktop counterparts. For instance, the Exynos 9820 utilizes an NPU that executes zero-skip operations to improve efficiency from 2.1 TOPS to 6.9 TOPS while handling 0\% and 75\% zero-weights networks, resulting in an energy efficiency increase from 3.6 TOPS/W to 11.5 TOPS/W \cite{song2019topsmobile}. This serves to confirm the viability of directing similar attacks towards on-device machine learning systems. Consequently, in our experimentation, we have opted to utilize mobile phones with distinct chips from various generations as our platform, as opposed to other on-device platforms.


\section{Methodology}

\subsection{Threat Model}\label{Threat Model}
Our threat model is following the extensive work of data poisoning and backdoor attack \cite{doan2021liraThreatModel1}\cite{gu2017badnetsThreatModel2}\cite{liu2017trojaningThreatModel3}\cite{yao2019latentThreatModel4}\cite{liao2018backdoorThreatModel5}.In our threat model, we consider the scenario where victim clients outsource their training process to third-party service providers who charge fees for providing server clusters with powerful computing capabilities. \par
The attacker aims to slow down on-device model inference and drain phone power. To achieve this objective, the attacker may hijack the training process or act as a man-in-the-middle to tamper with the examples and increase the average training cost. Meanwhile, the attacker has to guarantee no degradation in terms of model accuracy just in case the abnormal accuracy change in the attacker's unknown validation set will draw victims arouse. We assume that the poisoned model will pass the accuracy check from the victim's side, which means that it should perform well on the victim's validation set. Once the model has passed the accuracy check, it will be distributed to local mobile users, and the attacker cannot affect the localized models when they perform model inference.

\subsection{Attack Strategies}\label{Attack Strategies}

\begin{algorithm*}
\caption{Sponge Poisoning Attack via Training Process}\label{Attack Algorithm}
\begin{algorithmic}
\STATE \textbf{Input:} $\mathcal{D, P}$
\STATE \textbf{Output:} $w$
\end{algorithmic}
\begin{algorithmic}[1]
\STATE $w\gets$ \verb|random_init()|   \COMMENT{initialize model's weights}
\FOR{$i$ in $0,\dots,N$}
    \FOR{$(x,y)$ in $\mathcal{D}$}
        \IF {$(x,y)$ in $\mathcal{P}$} 
            \STATE $w(i+1) \gets w'(i)-\alpha [\nabla_{w} \mathcal{L}(x,y,w(i)) - \lambda \nabla_{w} E(x,w(i))]$    \COMMENT{update weights with sponge attack}
        \ELSE
            \STATE $w(i+1) \gets w'(i)-\alpha \nabla_{w} \mathcal{L}(x,y,w(i))$    \COMMENT{update weight without sponge}
        \ENDIF 
    \ENDFOR
\ENDFOR
\STATE \textbf{return} $w(N)$
\end{algorithmic}
\end{algorithm*}
\noindent In this work, to facilitate the characteristics of deployment for an on-device model and to perform sponge poisoning in such a system, we construct our experiment process in the following three folds:
\begin{itemize}
    \item We train model in the light-weight networks rather than a conventional network like ResNet and VGG to adapt the experiment in the mobile environment.
    \item We evaluate the model with various configurations of hyper-parameters to analyze their solo or joint impacts on the energy consumption in the PC end following the same metrics as Sponge Poisoning
    \item Subsequently, we deploy the model in actual devices to verify the accountability of such an attack.
\end{itemize}


\subsubsection{Sponge Training} \label{Sponge Training}
We use the same training algorithm as Sponge Poisoning Attack \cite{cina2022spongepoisoning} as follows: Given training dataset $\mathcal{D}$, we assume the attacker only has control of a portion p\% of it, represented by $\mathcal{P}$. As shown in Algorithm \ref{Attack Algorithm}, the attacker judges the current data source during the training process and applies the attack only for the subset $\mathcal{P}$. If the current data is not contained by $\mathcal{P}$, updates the model's weight with a normal approach, otherwise, the attacker applys the sponge poisoning attack. \par

When encountering samples from $\mathcal{P}$, we then adjust our update procedure to instruct the model towards more energy expense using \eqref{Weight update}. To be specific, we add a regularization term into the gradient update process where $E$ is the objective function that corresponds to the increase of model activation sparsity. By Maximizing this value, we can control the training towards more energy consumption. 


\begin{equation}
\min_w\sum_{(x,y)\in \mathcal{D}} \mathcal{L}(x,y,w) - \lambda \sum_{(x,y)\in \mathcal{P}} E(x,f,w) \label{Weight update}
\end{equation}

Details of the differentiable function $E$ are shown in \eqref{differentiable function}. Given that model $f$ having $K$ layers, along with input $x$ using parameters $w$, $\phi_{k}=f_{k}(x,w)$ denotes activations in the $k$-layer.\par

\begin{equation}
E(x,w) = \sum_{k=1}^{K} \hat{\ell_{0}} (\phi_{k}) \label{differentiable function}
\end{equation}

To increase energy consumption and latency by rising the model's activation, it is most objectively suitable to maximize the $\ell_{0}$ pseudo norm, which counts the number of non-zero elements of the model. Although the $\ell_{0}$ norm is a non-convex and discontinuous function for which optimization is NP-hard. Sponge Poisoning Attack is using the unbiased estimate of $\ell_{0}$ by formulation proposed in \cite{64de2011deconvolution}, shown as $\hat{\ell_{0}}$ in \eqref{Single Activation Density}.

\begin{equation}
\hat{\ell_{0}} (\phi_{k}) = \sum_{j=1}^{d_{k}} \frac{\phi_{k,j}^{2}}{\phi_{k,j}^{2} + \sigma}; \quad \phi_{k} \in \mathbb{R}^{d_{k}}, \sigma \in \mathbb{R} \label{Single Activation Density}
\end{equation}

\subsubsection{Hyper-parameter Selection}
To look for the empirical evidence of threats of sponge poisoning in real-world deployment, we aim to find out how the attack will impact the model with different combinations of hype-parameters in the task.
There are three hyper-parameters from Sponge Poisoning Attack:
\begin{itemize}
    \item $\mathcal{P}$: the portion of the dataset that applies attack formulation by percentage.
    \item $\lambda$: the Lagrangian penalty term that controls the strength of the attack.
    \item $\sigma$: the smaller the value is, the closer $\hat{\ell_{0}}$ to real $\ell_{0}$.
\end{itemize}
By configuring the main factors of sponge attack, we can have an analysis and basic understanding towards this area.



\section{Experimental Setup}\label{Experimental Setup}
\subsection{Datasets}
Our experiment is evaluated on CIFAR-10 \cite{krizhevsky2009cifar10}, which is the popular object recognition dataset. Both datasets contain 60K RGB images with a size of 32×32. Where the CIFAR-10 images are defined by 10 classes. We adopted a standard data practice scheme in our experiments: 50K examples are used for training and the remaining 10K are used for testing.\par
\subsection{Model Selection}
We apply our experiments on well-known compact network architectures that are suitable for on-device scenarios, including MobileNetV2 \cite{sandler2018mobilenetv2} and MobileNetV3 \cite{howard2019mobilenetv3}. Even they are consistent varieties at two stages, these two works are widely applied to mobile devices and tasks. For example,   mobilenetV2 is managed being delopyed to do facial recognition task in a Raspberry Pi\cite{Wang2021piv2} and MobileNetV3 further enhances its adaption ability to search for network architecture for different needs. Beside, the overall size of these two models are less than 10M which is ideal for our experiment in a lightweight case.

\subsection{Device and Baseline Setup}\label{Device Setup}
 Based on the discussion in section \ref{Mobile Phone Platforms}, we are choosing platforms that are more versatile for on-device scenarios with the following considerations: As of 2022, Qualcomm maintains the highest proportion of all mobile SoCs around the world with 44\% market share \cite{soc2020marketshar}. Qualcomm has also made deployments for autonomous vehicle scenarios. Because of its largest market share and diverse deployment scenario, we first select their latest processor Qualcomm Snapdragon 8Gen1 as our test platform. To illustrate the difference in terms of attack performance, we choose the Snapdragon 845 which is first generation of mobile soc with hradware-level acceleration techique as our auxilary verifciation platform.
 \begin{table*}[htbp]
\caption{MobileNetV3 Battery Historian Statistics}
\begin{center}
\begin{tabular}{|c|c|c|c|}
\hline
    \textbf{Model Setting} & \textbf{Battery Consumption (\%)} & \textbf{Middle Discharge Rate (\%)} & \textbf{Peak Discharge Rate (\%)} \\
\hline
    \multicolumn{4}{|c|}{Snapdragon 8Gen1}\\
\hline
    Unattacked & 16 & 35.96 & 39.96 \\
\hline
    1 & 17 & 39.96 & 44.95 \\
\hline
    10 & 20 & 44.95 & 51.37 \\
\hline
    20 & 21 & 44.95 & 51.37 \\
\hline
    \multicolumn{4}{|c|}{Snapdragon 845}\\
\hline
    Unattacked & 16 & 21.97 & 31.96 \\
\hline
    1 & 18 & 23.44 & 35.16 \\
\hline
    10 & 19 & 23.44 & 35.16 \\
\hline
    20 & 18 & 27.04 & 35.16 \\
\hline
\multicolumn{4}{l}{$^{\mathrm{a}}$The $\mathcal{P}$ and $\sigma$ of all selected attacked models are unified to 0.25 and 1e-06, respectively. The column "model" represents the difference of $\lambda$.}
\end{tabular}
\label{tab1}
\end{center}

\end{table*}
To conduct an empirical analysis of the energy consumption of a mobile device running a neural network, we simulate extreme conditions that require the network to run continuously for a long period. This enables us to observe any differences in energy consumption and potential risks that may arise when the device faces streaming data. To achieve this, we develop an Android App that continuously verifies the validation set on the mobile phone using a trained model with 100,000 samples.\par
To establish a baseline for the experiment, we used a clean trained model and recorded the runs of the test for 100,000 samples, which consumed 15\% of the device's battery. We then compared this to the energy consumption of the attacked model with the same number of runs. The device used in the experiment was equipped with Snapdragon 8Gen1 and had a battery capacity of 4600mAh. We needed 100 epochs using the unattacked model to drop the battery level from 100\% to 85\%, after which we recharged the device to 100\% and ran another 100 epochs with the attacked model. \par
Furthermore,To eliminate any potential factors that could cause extra energy consumption, we conduct the experiment in flight mode with all other programs closed and the screen brightness turns to the lowest possible level. Given that the Snapdragon 8Gen1 generates a significant amount of heat, most mobile device manufacturers have implemented temperature control strategies to prevent overheating. As such, it is challenging to bypass this limitation completely, and the device will inevitably be throttled due to overheating. We address this limitation by recording the data of the first ten epochs at the beginning of the test separately and treating the data of subsequent epochs as stable primary data. \par

\subsection{Evaluation Metrics}
One main metric is from Sponge Example \cite{shumailov2021spongeexample} where they use the ASIC simulator to measure the difference in energy ratio with and without the zero-skipping acceleration strategy. In our work, we adapt the simulator into our model to select the most efficient hyper-parameter combination with its energy ratio indicator. As the combination of this method triggers most energy consumption, we further deploy such a setting into the device to verify the accountability of this attack.

However, the accurate energy consumption measurement for Android devices required by the third party is too difficult to achieve, since Google has banned users from accessing several low-level APIs after 2018 for system stability reasons \cite{google2018api}. Therefore, our experiment for the third step will be held based on side data and one estimated energy consumption provided by Google:

\begin{itemize}

\begin{figure}[h]
        \centering
        \includegraphics[width=0.9\linewidth]{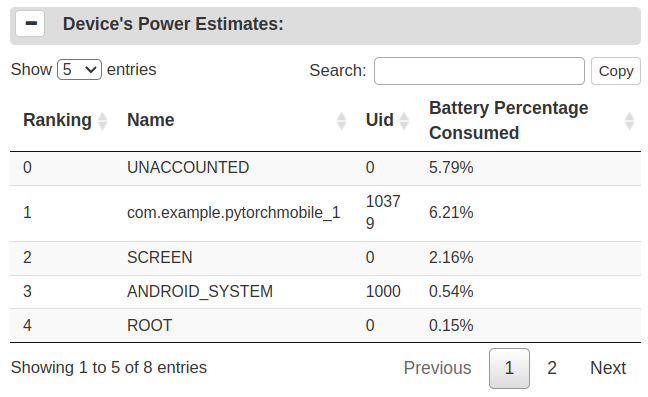}
        \caption{Google Battery Historian.}
        \Description{Screenshot of Google's BatteryHistorian Dashboard.}
        \label{BatteryHistorianFig}
\end{figure}

    \item {\textit{Battery Historian}}: The analysis visualization script tool officially provided by Google. Users can use the 'adb' command to collect battery usage data and logs from mobile phones. Its visual analysis includes the estimation of approximate power, as shown in Fig.~\ref{BatteryHistorianFig}.\par
    To be noticed Android can't directly map the power usage to each application. In the meantime, Android uses the power consumption of other major components to estimate the power consumption of a single application. For example, a component (such as SoC, screen, or 5G module) uses 100mAh of power in 20 minutes, and the program occupies this component for 10 minutes. It results in the power consumption of the program will be estimated to be 50mAh. Based on this principle, Android gives the estimated power consumption of a single program, which often has considerable errors. \par

\begin{figure*}[htb]
    \begin{minipage}[t]{0.3\textwidth}
        \centering
        \begin{tikzpicture}[thick,scale=0.65, every node/.style={transform shape}]
            \begin{axis}[
                ylabel={Energy Ration (\%)},
                xmin=0, xmax=13,
                ymin=88, ymax=100,
                xtick={0,1,2,4,5,6,7,9,12},
                ytick={88,89,90,91,94,95,96,97},
                legend pos=north west,
                ymajorgrids=true,
                grid style=dashed,
            ]
            
            \addplot[
                color=blue,
                mark=square,
                ]
                coordinates {
                (1,89.63316679)(2,89.73343968)(4,89.82406855)(5,90.14712572)(6,90.32152891)(7,90.160954)(9,89.8194015)
                };
                \legend{$\mathcal{P}$:0.05 \& $\lambda$:1}
            \addplot [
            dashed,
            domain=0:13, 
            color=black,
            ]
            {88.4};
        
            \addplot[
                color=red,
                mark=triangle,
                ]
                coordinates {
                (4,96.191293)(5,96.41491771)(6,96.39727473)(7,96.5749383)(8,96.64903879)(9,97.29844928)(12,94.49065328)
                };
                \addlegendentry{$\mathcal{P}$:0.35 \& $\lambda$:10}
        
            \end{axis}
            \end{tikzpicture}
            
        \begin{tikzpicture}[thick,scale=0.65, every node/.style={transform shape}]
        \begin{axis}[
            xlabel={$\sigma$ by 1e-$x$},
            ylabel={Validation Accuracy (\%)},
            xmin=0, xmax=13,
            ymin=75, ymax=85,
            xtick={0,1,2,4,5,6,7,9,12},
            ytick={},
            legend pos=north west,
            ymajorgrids=true,
            grid style=dashed,
        ]
        
        \addplot[
            color=blue,
            mark=square,
            ]
            coordinates {
            (1,78.99)(2,80.4)(4,80.24)(5,81.01)(6,81)(7,82.66)(9,82.13)
            };
            \legend{$\mathcal{P}$:0.05 \& $\lambda$:1}
        \addplot [
        dashed,
        domain=0:13, 
        color=black,
        ]
        {81.23};
        
        \addplot[
            color=red,
            mark=triangle,
            ]
            coordinates {
            (4,77.75)(5,81.11)(6,82)(7,81.68)(8,83)(9,72.89)(12,32)
            };
            \addlegendentry{$\mathcal{P}$:0.35 \& $\lambda$:10}
        
        \end{axis}
    \end{tikzpicture}
            
        \caption{Energy Ratio by $\sigma$ for MobileNetV3 while $\mathcal{P}$:0.05 \& $\lambda$:1, along with the same model's validation accuracy. [The black dashed line stands for the baseline of unattacked model]}
        \label{fig2}
    \end{minipage}
    \hfill
    \begin{minipage}[t]{0.3\textwidth}
        \centering
        \begin{tikzpicture}[thick,scale=0.65, every node/.style={transform shape}]
        \begin{axis}[
            ylabel={Energy Ration (\%)},
            xmin=20, xmax=80,
            ymin=95, ymax=100,
            xtick={20,25,35,45,55,65,75,80},
            ytick={},
            legend pos=north west,
            ymajorgrids=true,
            grid style=dashed,
        ]
        
        \addplot[
            color=blue,
            mark=square,
            ]
            coordinates {
            (25,98.39375615)(35,98.21808934)(45,98.34336042)(55,98.25693965)(65,98.37642908)(75,98.25376272)
            };
            \legend{$\sigma$:1e-06 \& $\lambda$:20}
        \addplot [
        dashed,
        domain=0:13, 
        color=black,
        ]
        {88.4};
    
        \end{axis}
        \end{tikzpicture}
        
        \begin{tikzpicture}[thick,scale=0.65, every node/.style={transform shape}]
        \begin{axis}[
            xlabel={$\mathcal{P}$ by \%},
            ylabel={Validation Accuracy (\%)},
            xmin=20, xmax=80,
            ymin=75, ymax=85,
            xtick={20,25,35,45,55,65,75,80},
            ytick={},
            legend pos=north west,
            ymajorgrids=true,
            grid style=dashed,
        ]
        
        \addplot[
            color=blue,
            mark=square,
            ]
            coordinates {
            (25,79.48)(35,81)(45,79.62)(55,81.45)(65,80.64)(75,80.24)
            };
            \legend{$\sigma$:1e-06 \& $\lambda$:20}
        \addplot [
        dashed,
        domain=0:13, 
        color=black,
        ]
        {81.23};
        
        \end{axis}
    \end{tikzpicture}
            
        \caption{Energy Ratio by $\mathcal{P}$ for MobileNetV3 while $\sigma$:1e-06 \& $\lambda$:20, along with the same model's validation accuracy.}
        \label{fig3}
    \end{minipage}  
    \hfill
    \begin{minipage}[t]{0.3\textwidth}
        \centering
        \begin{tikzpicture}[thick,scale=0.65, every node/.style={transform shape}]
        \begin{axis}[
            ylabel={Energy Ration (\%)},
            xmin=0, xmax=31,
            ymin=85, ymax=100,
            xtick={1,5,10,15,20,25,30},
            ytick={},
            legend pos=north west,
            ymajorgrids=true,
            grid style=dashed,
        ]
        
        \addplot[
            color=blue,
            mark=square,
            ]
            coordinates {
            (1,89.47765231)(5,93.86585951)(10,96.19657397)(15,97.80257344)(20,98.39375615)(25,98.88554811)(30,98.80749583)
            };
            \legend{$\sigma$:1e-06 \& $\mathcal{P}$:0.25}
        \addplot [
        dashed,
        domain=0:31, 
        color=black,
        ]
        {88.4};
    
        \end{axis}
        \end{tikzpicture}
        
        \begin{tikzpicture}[thick,scale=0.65, every node/.style={transform shape}]
        \begin{axis}[
            xlabel={$\lambda$ by \%},
            ylabel={Validation Accuracy (\%)},
            xmin=0, xmax=31,
            ymin=75, ymax=85,
            xtick={1,5,10,15,20,25,30},
            ytick={},
            legend pos=north west,
            ymajorgrids=true,
            grid style=dashed,
        ]
        
        \addplot[
            color=blue,
            mark=square,
            ]
            coordinates {
            (1,81.69)(5,81.6)(10,81.46)(15,81.07)(20,79.48)(25,78.73)(30,77.81)
            };
            \legend{$\sigma$:1e-06 \& $\mathcal{P}$:0.25}
        \addplot [
        dashed,
        domain=0:31, 
        color=black,
        ]
        {81.23};
        
        \end{axis}
    \end{tikzpicture}
        
        \caption{Energy Ratio by $\lambda$ for MobileNetV3 while $\sigma$:1e-06 \& $\mathcal{P}$:0.25, along with the same model's validation accuracy. [The black dashed line stands for the baseline of unattacked model]}
        \label{fig4}
    \end{minipage}  
\end{figure*}

\begin{table*}[htbp]
\caption{MobileNetV3 Software-based Data Collection}
\begin{center}
\begin{tabular}{|c|c|c|c|c|c|c|c|c|}
\hline
    \textbf{Model} & \multicolumn{4}{|c|}{\textbf{Stable Epochs}} & \multicolumn{4}{|c|}{\textbf{First 10 Epochs}} \\
    \cline{2-9}
    \textbf{Setting} & \textbf{Time (ms)} & \textbf{Current (mA)} & \textbf{Voltage (mV)} & \textbf{Watt (mW)} & \textbf{Time (ms)} & \textbf{Current (mA)} & \textbf{Voltage (mV)} & \textbf{Watt (mW)}\\
\hline
    \multicolumn{9}{|c|}{Snapdragon 8Gen1}\\
\hline
    Unattacked & 2229189 & 1496 & 4224 & 6324950 & 194838 & 2156 & 4320 & 9318690 \\
\hline
    1 & 2026617 & 1657 & 4207 & 6976045 & 194739 & 2216 & 4294 & 9518596\\
\hline
    10 & 1933137 & 1688 & 4131 & 6977514 & 186602 & 2178 & 4249 & 8259166\\
\hline
    20 & 2032738 & 1697 & 4168 & 7080550 & 196894 & 2239 & 4269 & 9564242\\
\hline
    \multicolumn{9}{|c|}{Snapdragon 845}\\
\hline
    Unattacked & 1442148 & 838 & 4048 & 3395663 & 399311 & 1090 & 4032 & 4399836 \\
\hline
    1 & 1438346 & 850 & 4026 & 3426386 & 396349 & 1096 & 4004 & 4383840\\
\hline
    10 & 1438437 & 851 & 4033 & 3435771 & 392823 & 1128 & 3998 & 4510960\\
\hline
    20 & 1416057 & 855 & 4025 & 3445101 & 387300 & 1150 & 3979 & 4578618\\
\hline
\multicolumn{9}{l}{$^{\mathrm{a}}$The $\mathcal{P}$ and $\sigma$ of all selected attacked models are unified to 0.25 and 1e-06, respectively. The column "model" represents the difference of $\lambda$.}
\end{tabular}
\label{tab2}
\end{center}
\end{table*}
    
    In this case, if the total estimated power of all components and programs is less than the actual power consumption obtained from the battery, the system will classify it as "unaccounted". Nevertheless, It seems that the "unaccounted" section denoted by this approach should be attributed to the effect of running programs as well in our settings, the details will be explained in the section \ref{Analysis Discussion}.
    
    \item {\textit{Discharge Rate}}: The Discharge Rate is an important metric provided by Google Battery Historian for analyzing power consumption patterns. It is reported in intervals of 100 seconds and indicates the overall power consumption trend for the corresponding segment, such as 30\% per hour. The Discharge Rate is estimated and provided in stepped-levels, such as 36/40/45/52 per hour, which may vary for different mobile device models. \par
    In our experiment, we set the baseline settings to operate the device for approximately 30 to 40 minutes, and recorded the estimated Discharge Rate for a dozen segments. We primarily focused on the median and maximum values of the Discharge Rate to evaluate the power consumption behavior of the device.

    \item {\textit{Run-Time Statistics}}: Since the Battery Historian's data can not accurately reveal the data consumption, we also collect real-time information about the operation of the equipment. Refer to \cite{tarkoma2014smartphoneenergyconsumption}, we are applying a software-based measurement based on Android API "BatteryManager". We are allowed to collect the present voltage and current statistics and thus calculate the current watt. However, as the property of software-based measurement, it relies on the internal code of the running program to obtain related data\cite{perrucci2011energysurvey}. For the above reasons, according to our settings in \ref{Device Setup}, each epoch contains 10K data from the CIFAR dataset. We collect voltage and current every 100 data in each epoch, which delivers 100 sets of data in each epoch. Our analysis is based on the summed-up data for the entire test process. \par
    It is necessary to note that the voltage and current information we collect is actually from the battery, not the phone's motherboard. This is also a common built-in measurement approach from most devices. The battery level estimation of mobile phones is usually performed by a special "fuel gauge" chip. Different phone models may use different chips which depend on their manufacturer. However, most chips in modern devices can often measure both voltage and current, and sometimes battery temperature. Such chips could potentially provide a more precise estimate of battery information, thus the energy information. This is also the source of our data.
    
    \item {\textit{Direct Battery Level}}: As the most intuitive data source, we also collected the difference in the percentage of battery displayed after each experiment. Since we start the experiment after fully charging the battery every time, the controlled data is comparable in the experiment setting. Its intuitive gap also guides our subsequent analysis.
    
    \item {\textit{Latency of App}}: Our test app will first read the dataset, and then run uninterruptedly a specific number of epochs. Java's built-in method is employed to record the running time of each epoch and the overall running time.
\end{itemize}

\section{Analysis of Results and Discussion} \label{Analysis Discussion}

\subsection{MobileNetV3 Results}

\begin{figure*}[htb]
    \begin{minipage}[t]{0.45\textwidth}
        \centering
        \begin{tikzpicture}[thick,scale=0.75, every node/.style={transform shape}]
            \begin{axis}[
                ylabel={Energy Ration (\%)},
                xmin=0, xmax=10,
                ymin=75, ymax=100,
                xtick={0,1,2,4,5,6,7,9,10},
                ytick={},
                legend pos=north west,
                ymajorgrids=true,
                grid style=dashed,
            ]
            
            \addplot[
                color=blue,
                mark=square,
                ]
                coordinates {
                (1,81.45278096)(2,87.80961037)(4,92.0343399)(5,93.75475645)(6,93.42549443)(7,94.33276057)(9,94.04639006)
                };
                \legend{$\mathcal{P}$:0.25 \& $\lambda$:10}
            \addplot [
            dashed,
            domain=0:13, 
            color=black,
            ]
            {76.1674881};
        
            \end{axis}
            \end{tikzpicture}
            
            \begin{tikzpicture}[thick,scale=0.75, every node/.style={transform shape}]
            \begin{axis}[
                xlabel={$\sigma$ by 1e-$x$},
                ylabel={Validation Accuracy (\%)},
                xmin=0, xmax=10,
                ymin=60, ymax=85,
                xtick={0,1,2,4,5,6,7,9,10},
                ytick={},
                legend pos=north west,
                ymajorgrids=true,
                grid style=dashed,
            ]
            
            \addplot[
                color=blue,
                mark=square,
                ]
                coordinates {
                (1,63.24)(2,75)(4,79.7)(5,78.33)(6,80.82)(7,77.1)(9,74.54)
                };
                \legend{$\mathcal{P}$:0.25 \& $\lambda$:1}
            \addplot [
            dashed,
            domain=0:13, 
            color=black,
            ]
            {81.02};
            
            \end{axis}
        \end{tikzpicture}
            
        \caption{Energy Ratio by $\sigma$ for MobileNetV2 while $\mathcal{P}$:0.25 \& $\lambda$:10, along with the same model's validation accuracy. [The black dashed line stands for the baseline of unattacked model]}
        \label{fig5}
    \end{minipage}
    \hspace{2mm}
    \begin{minipage}[t]{0.45\textwidth}
        \centering
        \begin{tikzpicture}[thick,scale=0.75, every node/.style={transform shape}]
        \begin{axis}[
            ylabel={Energy Ration (\%)},
            xmin=0, xmax=31,
            ymin=75, ymax=100,
            xtick={1,5,10,15,20,25,30},
            ytick={},
            legend pos=north west,
            ymajorgrids=true,
            grid style=dashed,
        ]
        
        \addplot[
            color=blue,
            mark=square,
            ]
            coordinates {
            (1,80.13997078)(5,89.67282176)(10,93.42549443)(15,94.23108101)(20,95.81977725)(25,95.34759521)(30,95.38792372)
            };
            \legend{$\mathcal{P}$:0.25 \& $\sigma$:1e-06}
        \addplot [
        dashed,
        domain=0:31, 
        color=black,
        ]
        {76.1674881};
    
        \end{axis}
        \end{tikzpicture}
        
        \begin{tikzpicture}[thick,scale=0.75, every node/.style={transform shape}]
        \begin{axis}[
            xlabel={$\lambda$},
            ylabel={Validation Accuracy (\%)},
            xmin=0, xmax=31,
            ymin=45, ymax=85,
            xtick={1,5,10,15,20,25,30},
            ytick={45,60,65,75,80,85},
            legend pos=south west,
            ymajorgrids=true,
            grid style=dashed,
        ]
        
        \addplot[
            color=blue,
            mark=square,
            ]
            coordinates {
            (1,80.12)(5,81.37)(10,80.82)(15,64.69)(20,61.62)(25,49.37)(30,60.89)
            };
            \legend{$\mathcal{P}$:0.25 \& $\sigma$:1e-06}
        \addplot [
        dashed,
        domain=0:31, 
        color=black,
        ]
        {81.02};
        
        \end{axis}
    \end{tikzpicture}
        
        \caption{Energy Ratio by $\lambda$ for MobileNetV2 while $\mathcal{P}$:0.25 \& $\sigma$:1e-06, along with the same model's validation accuracy. [The black dashed line stands for the baseline of unattacked model]}
        \label{fig6}
    \end{minipage}  
\end{figure*}

\begin{table*}[htbp]
\caption{MobileNetV2 Battery Historian Statistics}
\begin{center}
\begin{tabular}{|c|c|c|c|}
\hline
    \textbf{Model Setting} & \textbf{Battery Consumption (\%)} & \textbf{Middle Discharge Rate (\%)} & \textbf{Peak Discharge Rate (\%)} \\
\hline
    \multicolumn{4}{|c|}{Snapdragon 8Gen1}\\
\hline
    Unattacked & 16 & 35.96 & 44.95 \\
\hline
    Attacked & 20 & 44.95 & 50.96 \\
\hline
    \multicolumn{4}{|c|}{Snapdragon 845}\\
\hline
    Unattacked & 14 & 26.50 & 39.06 \\
\hline
    Attacked & 16 & 23.28 & 39.06 \\
\hline
\multicolumn{4}{l}{$\mathcal{P}=0.25; \sigma=1e-06; \lambda=10 $ for the selected attacked model.}
\end{tabular}
\label{tab3}
\end{center}

\end{table*}

\begin{table*}[htbp]
\caption{MobileNetV2 Software-based Data Collection}
\begin{center}
\begin{tabular}{|c|c|c|c|c|c|c|c|c|}
\hline
    \textbf{Model} & \multicolumn{4}{|c|}{\textbf{Stable Epochs}} & \multicolumn{4}{|c|}{\textbf{First 10 Epochs}} \\
    \cline{2-9}
    \textbf{Setting} & \textbf{Time (ms)} & \textbf{Current (mA)} & \textbf{Voltage (mV)} & \textbf{Watt (mW)} & \textbf{Time (ms)} & \textbf{Current (mA)} & \textbf{Voltage (mV)} & \textbf{Watt (mW)}\\
\hline
    \multicolumn{9}{|c|}{Snapdragon 8Gen1}\\
\hline
    Unattacked & 1619946 & 1657 & 4212 & 6987581 & 149972 & 2165 & 4316 & 9348716 \\
\hline
    Attacked & 1625678 & 1663 & 4142 & 6893553 & 156419 & 2175 & 4295 & 9347685\\
\hline
    \multicolumn{9}{|c|}{Snapdragon 845}\\
\hline
    Unattacked & 1156909 & 835 & 4058 & 3392387 & 313546 & 1150 & 4011 & 4616918 \\
\hline
    Attacked & 1208471 & 805 & 4068 & 3278183 & 320854 & 1101 & 4022 & 4429077\\
\hline
\multicolumn{9}{l}{$\mathcal{P}=0.25; \sigma=1e-06; \lambda=10 $ for the selected attacked model.}
\end{tabular}
\label{tab4}
\end{center}
\end{table*}

\subsubsection{Impact of $\sigma$}
We first conduct an attack training against MobileNetV3. As mentioned in the section \ref{Attack Strategies}, we test different hyper-parameters and select the most potential result for the next step of verification. At the same time, it is also worthwhile to provide a certain preset basis for subsequent experiments on other network architectures. We first perform hyper-parameter sensitivity on MobileNetV3.\par
We firstly inspect the impact for different value choices of $\sigma$, Fig.~\ref{fig2} shows the variation of energy ratio by training attacked models with different $\sigma$ with fixed $\mathcal{P}$ and $\lambda$. There are two sets of $\mathcal{P}$ and $\lambda$ chosen which represent simulated extreme attack scenarios. Based on our assumption, the attacker is supposed to manipulate the process as less as possible, thus we choose $\mathcal{P}$ = 0.05 which means only 5\% of the dataset will be attacked during the training process. We also set $\lambda$ to a relatively low level that $\lambda$ = 1.  In the second set of data, $\mathcal{P}$ and $\lambda$ are approaching the limit envisaged by the threat model. That almost one-third of the dataset will be compromised, and $\lambda$ is also set to a value stable enough to observe the difference. However, it appears that the hyper-parameter $\mathcal{P}$'s impact is limited after a certain percentage from the figure.\par
Our observation shows that too large or small values of $\sigma$ are both likely to make the result worse in terms of accuracy and estimated energy ratio. The higher the value of $\sigma$ is, the lower the approximation of $\hat{\ell_{0}}$ will be. It makes the attack result less obvious. On the contrary, with the extreme well-fit approximation to $\ell_{0}$, the $\hat{\ell_{0}}$ is not sufficiently smooth to facilitate attack formulation. According to the results obtained by our test, combining the comprehensive consideration of attack effect and accuracy factors,  the learning rate for the model around 1e-06 is more suitable for MobileNetV3. Thus, the following experiments take $\sigma=1e-06$ as the default with $\lambda$ of 20. Comparing its stable phase data with the set with $\lambda$ value of 1, the delay is greater when the power consumption is higher. The first 10 epochs' readings further magnify and affirm this conclusion. On the contrary, it is difficult to deliver an obvious and positive conclusion for Snapdragon 845's data in this table. Referring to the observation results in Table.~\ref{tab1} from previous paragraphs, we may not be able to observe obvious differences due to the characteristics of the Snapdragon 845.

\subsection{MobileNetV2 Results}
Following the same experiment process of MObileNetV3, to investigate the impact of $\sigma$ on training, we first start with fixed $\mathcal{P}$=0.25 which is optimal for the last model. As shown in Fig.~\ref{fig5}, we can see as $lambda$ goes upwards, the energy ratios raise in an inclined trend. It reaches stable with an energy ratio around 98\% in the $lambda$ over 25\%. The validation accuracy declines in a consistent way when $lambda$ goes larger. To be noticed, with small $lambda$ before 10\%, validation accuracy surprisingly outperforms the unattacked model. With small $lambda$, one reasonable guess is that it introduced noise to prevent overfitting.\par
Based on 2 plots in Fig.~\ref{fig5}, we use $lambda$ 10 as our optimal value. Then we further implement the experiment to choose $\mathcal{P}$. According to Fig.~\ref{fig6}, the optimal learning rate is around 1e-6 which is the highest in validation accuracy with an enery ratio of around 94\%. \par
 


As experiments show consistency in terms of mobile platform test results along with the estimated result on the PC end, empirical results on the device present a similar curve as in the server. It verifies our concern in the mobile platform with no extra hardware accelerators. We further conduct an experiment to address the feasibility of this attack to different processors to show the most potential attack case. Attacks on these two processors do rise the energy consumption in models. For 8Gen1 and 845 there is 20\%  and 14\% extra consumption rising compared with the unattacked model. From data collected from the PC end, there is little difference to draw the conclusion of relations with performance between attacked and unattacked models. From the reading of 10 runs of work, the peak data shows that there is a longer ruining time at high power consumption. 
The test for the Snapdragon 845, however, does not show a significant gap. Neither the Battery Historian results nor the software-based collection readings, we can not deliver a firm conclusion on the attack effect. It also coincides with the experimental results of MobileNetV3. As we discussed before, Snapdragon 845 has inherent deficiencies in deep learning hardware acceleration. Besides, MobileNetV2 itself is more power-hungry than MobileNetV3, which makes it difficult to even observe the difference in discharge rate. 

\subsection{Experiment Discussion}
To summarize, the experiment from these two processors and models, we find that for Snapdragon 8Gen1, the energy ratio incline goes as expected for optimal attack configuration both in MobileNetV2 and MobileNetV3, the test on the PC end and the on-device end is consistent to show the realistic concerns of the potential for sponge poisoning in the real world. For older processor 845 which has less modern AI accelerators in contrast does not show enough sensitivity towards sponge poisoning.  Without compact hardware accelerator, it can be full computation power involved for SnapDragon 845.

\section{Conclusion and Future Work}
In this paper, we conduct an extensive exploration based on Sponge Poisoning Attack \cite{cina2022spongepoisoning} in the mobile setting which is energy-intensive and memory-intensive. The result shows the feasibility of this kind of attack using a crafted example with the objective of increasing power consumption and latency while maintaining the prediction performance. We have demonstrated the disparate impact of factors towards two generations of processors and show the attack can effectively deteriorate the state-of-art commercial processor embedded accelerators to cause energy rise. As the processors from the first generation are little affected by the attack, the mechanism level flaw inside hardware accelerators should gather more attention to avoid providing more attack surface for attackers.\par
Studying the energy consumption-related attacks on mobile devices can help make on-device machine learning more robust in the real-world running scenario. Additional processors with various mobile accelerators can be added in the future to extend the investigation in this field. This work merely considers the sponge poisoning with a specific method, other measurement techniques, networks method and metrics can be explored as future work.

\begin{acks}
This work was supported in part by the National Key R\&D Program of China (No. 2021YFB2900100), and the National Natural Science Foundation of China (No. 62002294, U22B2022).
\end{acks}

\bibliographystyle{ACM-Reference-Format}
\bibliography{main}

\appendix

\end{document}